\documentclass[10pt]{article}
\usepackage{graphicx}
\usepackage{subcaption}

\def\Title#1{\begin{center} {\Large #1 } \end{center}}
\def\Author#1{\begin{center}{ \sc #1} \end{center}}
\def\Address#1{\begin{center}{ \it #1} \end{center}}

\newcommand\pubblock{\rightline{\begin{tabular}{l} Proceedings of the Fifth Annual LHCP\\ \\
         \pubdate  \end{tabular}}}

\newenvironment{Abstract}{\begin{quotation} \begin{center} 
             \large ABSTRACT \end{center}\bigskip 
      \begin{center}\begin{large}}{\end{large}\end{center} \end{quotation}}

\newenvironment{Presented}{\begin{quotation} \begin{center} 
             PRESENTED AT\end{center}\bigskip 
      \begin{center}\begin{large}}{\end{large}\end{center} \end{quotation}}

\def\beq{\begin{equation}}
\def\eeq#1{\label{#1}\end{equation}}
\def\eeqn{\end{equation}}

\def\beqa{\begin{eqnarray}}
\def\eeqa#1{\label{#1}\end{eqnarray}}
\def\eeqan{\end{eqnarray}}

\let\bar=\overbar

\def\Dslash{\not{\hbox{\kern-4pt $D$}}}
\def\dslash{\not{\hbox{\kern-2pt $\del$}}}

\def\msb{{\bar{\ssstyle M \kern -1pt S}}}

\usepackage{color}

\newcommand\pubdate{\today}

\def\affiliation{
On behalf of the ATLAS Experiment, \\
DESY, Hamburg, Germany}

\begin{document}

\large
\begin{titlepage}
\pubblock

\vfill
\Title{  Simulation for the ATLAS Upgrade Strip Tracker  }
\vfill

\Author{ Jike Wang  }
\Address{\affiliation}
\vfill
\begin{Abstract}
ATLAS is making extensive efforts towards preparing a detector upgrade for the
High luminosity operations of the LHC (HL-LHC~\cite{hllhc}), which will commence operation
in about 10 years. The current ATLAS Inner Detector will be replaced by a
all-silicon tracker (comprising an inner Pixel tracker and outer Strip
tracker). The software currently used for the new silicon tracker is broadly
inherited from that used for the LHC Run-1 and Run-2, but many new developments
have been made to better fulfill the future detector and operation
requirements. One aspect in particular which will be highlighted is the
simulation software for the Strip tracker. The available geometry description
software (including the detailed description for all the sensitive elements,
the services, etc.) did not allow for accurate modelling of the planned
detector design. A range of sensors/layouts for the Strip tracker are being
considered and must be studied in detailed simulations in order to assess the
performance and ascertain that requirements are met. For this, highly
flexibility geometry building is required from the simulation software. A new
Xml-based detector description framework has been developed to meet the
aforementioned challenges. We will present the design of the framework and its
validation results.
\end{Abstract}

\vfill

\begin{Presented}
The Fifth Annual Conference\\
 on Large Hadron Collider Physics \\
Shanghai Jiao Tong University, Shanghai, China\\ 
May 15-20, 2017
\end{Presented}
\vfill
\end{titlepage}
\def\thefootnote{\fnsymbol{footnote}}
\setcounter{footnote}{0}
%

\normalsize 


\section{Studies for a next generation tracking detector}
An upgrade of the LHC is planned for around 2024 which will provide very high collision rates, 
enhancing the possibilities for new physics measurements, but also posing very significant experimental challenges. 
The current ATLAS Inner Detector (ID) will not be suitable for operation in such an environment, 
and so must be replaced; an all-silicon tracker is considered to be the best solution for such conditions. 
To arrive at the optimum design, detailed simulation studies are needed to predict the tracking performance of candidate layouts.

\section{High-Luminosity LHC}
The HL-LHC will operate with the following conditions: a proton-proton centre-of-mass energy $\sqrt{s}=14$~TeV; 
a luminosity of $L=5\times 10^{34}cm^{-2}s^{-1}$. 
Assuming luminosity leveling, this corresponds to 140-200 pile-up collisions per bunch crossing.
Such high pile-up will pose significant challenges to the detector, and to data analysis.

Figure \ref{fig:pileups} shows an event display of one event with up to 140 collisions (The plot is produced using the Atlantis software). 
The colored lines are the charged tracks in the ITk; the points are the vertices.  
We could see the very busy situation in the detector with the high pile-ups.

The schematic layout of the ITk for the HL-LHC is shown in Figure \ref{fig:layout}, 
here only one quadrant and only active detector elements are shown. 
The horizontal axis is the axis along the beam line with zero being the interaction point.
 The vertical axis is the radius measured from the interaction point. 
The outer radius is set by the bore of the solenoid. The outer blue color structure is the Strip tracker. 
 

\begin{minipage}[c]{.5\textwidth}
\includegraphics[height=0.25\textheight,width=1.\textwidth]{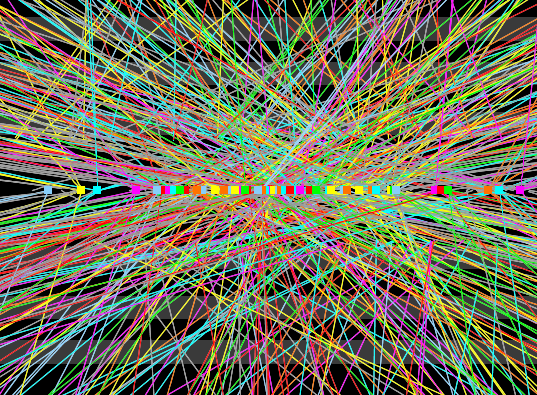}
\captionof{figure}{event display of a pile-up event}
\label{fig:pileups}
\end{minipage}
\hspace{0.2cm}
\begin{minipage}[c]{.5\textwidth}
\includegraphics[height=0.3\textheight,width=1.\textwidth]{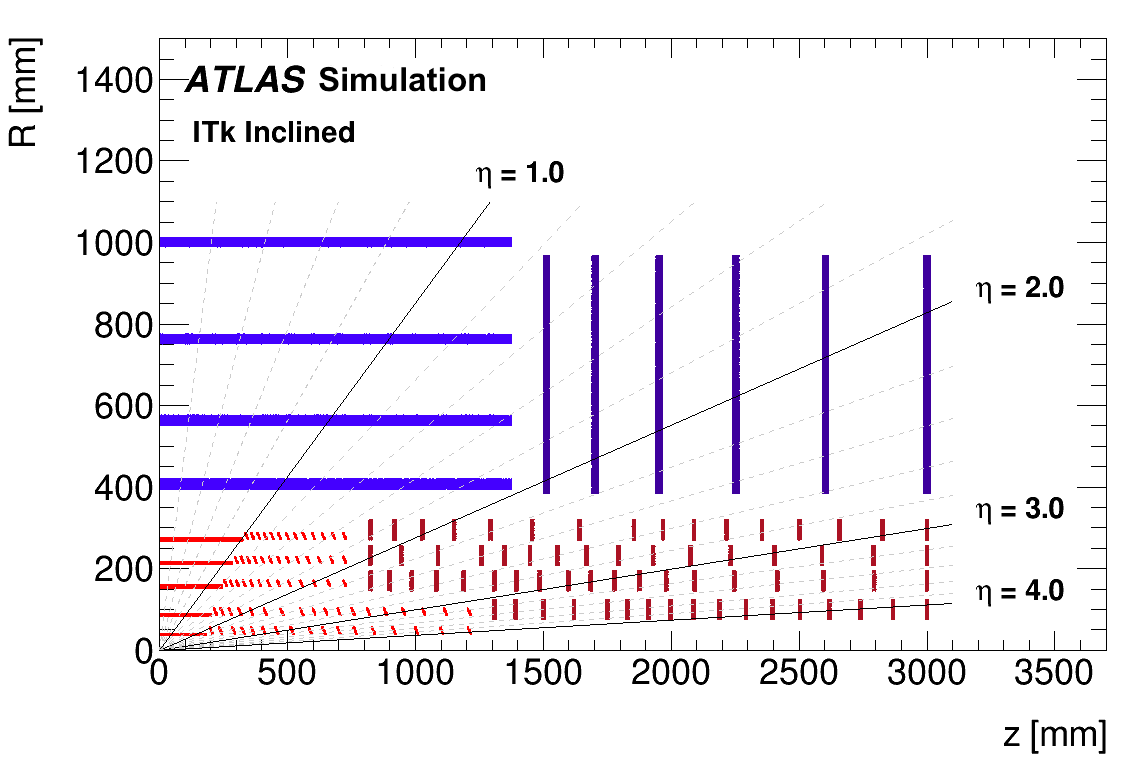}
\captionof{figure}{$R-Z$ view of the ITk}
\label{fig:layout}
\end{minipage}

\section{Building the Strip tracker using Xml}

For the current ATLAS Inner Detector, the geometry is defined directly in C++ code, 
with the dimensions and quantities stored in the text files.  
This way is very difficult to maintain and very hard to extend to new layouts. 
For the upgrade Strip tracker design, the Xml language is used. 
Xml is very well suited to geometry description: it can store numbers (such as dimensions, as well as their meaning), 
can also handle hierarchical structure, etc. The flow of this infrastructure is shown in Figure \ref{fig:flow}. 
Figure \ref{fig:ITk} shows a visualization of the ITk as implemented in the simulation framework; 
from inner to outer, there are the Pixel and Strip tracker respectively.

\begin{minipage}[c]{.5\textwidth}
\includegraphics[width=1.0\textwidth,height=0.23\textheight]{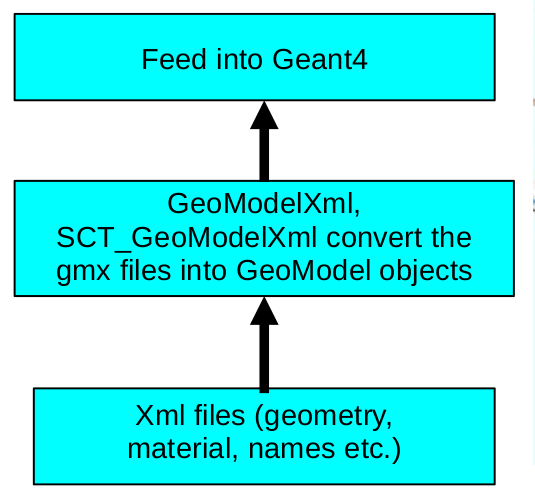}
\captionof{figure}{The flow chart of building the Strip tracker}
\label{fig:flow}
\end{minipage}
\hspace{0.2cm}
\begin{minipage}[c]{.5\textwidth}
\includegraphics[width=1.0\textwidth]{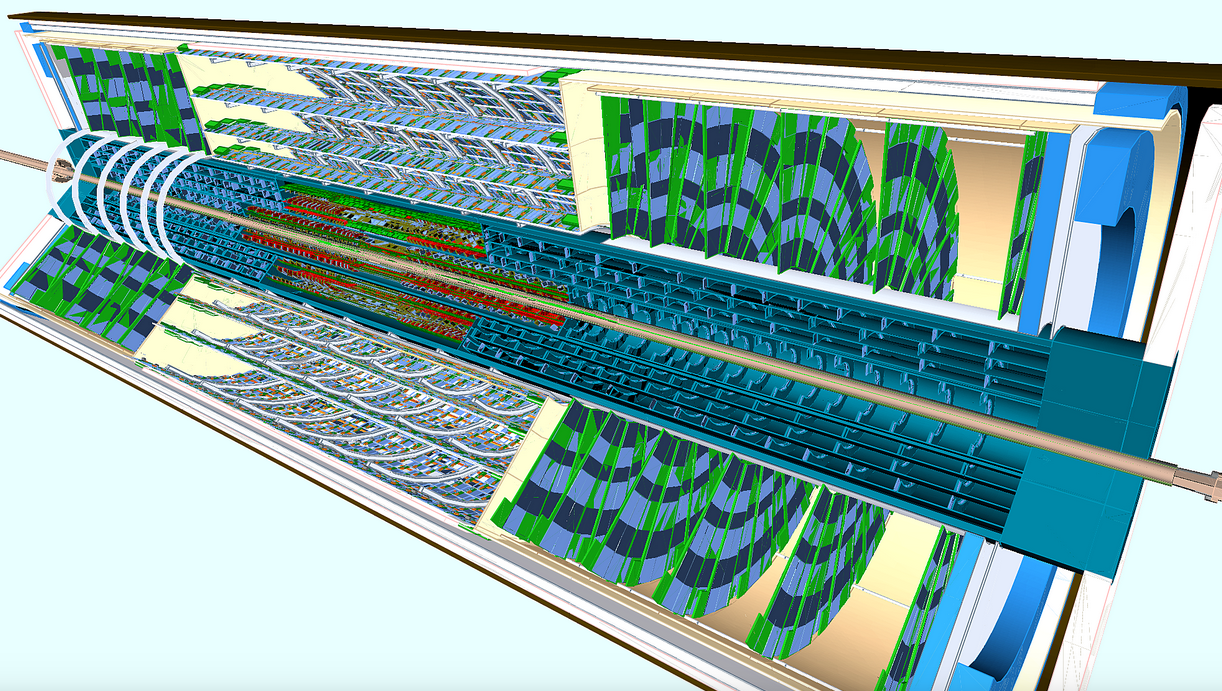}
\captionof{figure}{Visualization of the ITk as implemented in the simulation framework}
\label{fig:ITk}
\end{minipage}

\section{Design Considerations}

The following factors are amongst those taken into account when designing a detector for HL-LHC conditions:
\begin{itemize}
\item
Granularity should be sufficient to provide the necessary impact parameter resolution and allow good separation of tracks and vertices
\item
Material within the tracking acceptance should be kept to a minimum in order 
to reduce loss of resolution and efficiency through multiple-scattering and brems-strahlung. 
The material map of the ITk is shown in Figure \ref{fig:X0}. 
\item
Occupancy (the distribution as shown in Figure \ref{fig:Occup}) should be kept down to an acceptable level; 
this drives the decisions such as the length of the strips in the Strip detector
\item
The number of hit points must be sufficient to discriminate against fake tracks from combinatorics while maintaining a high tracking efficiency 
\end{itemize}

\begin{minipage}[c]{.495\textwidth}
\includegraphics[height=0.3\textheight,width=1.0\textwidth]{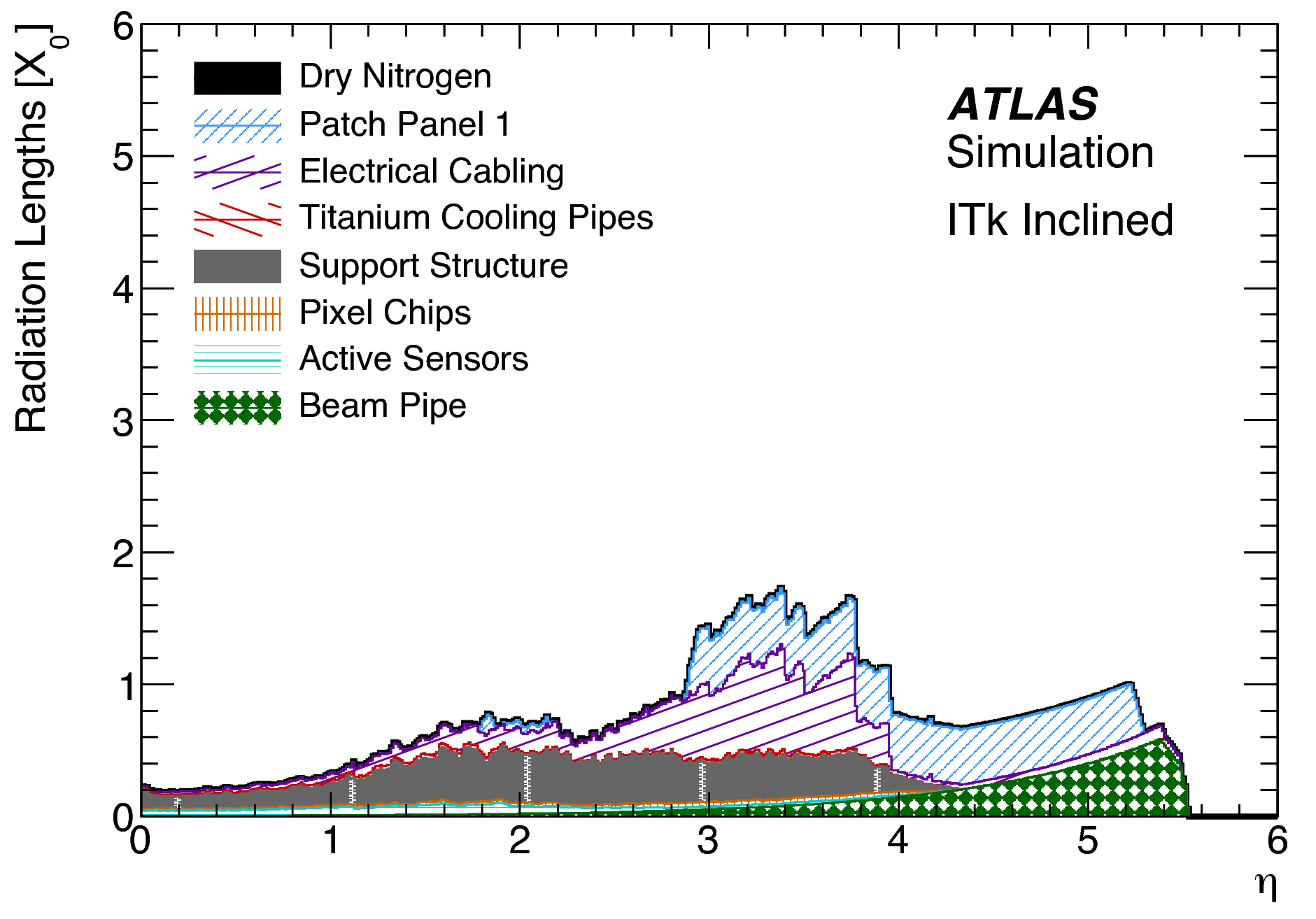}
\captionof{figure}{The material radiation length $X_0$ versus $\eta$ for the ITk}
\label{fig:X0}
\end{minipage}
\hspace{0.2cm}
\begin{minipage}[c]{.495\textwidth}
\includegraphics[height=0.3\textheight,width=1.0\textwidth]{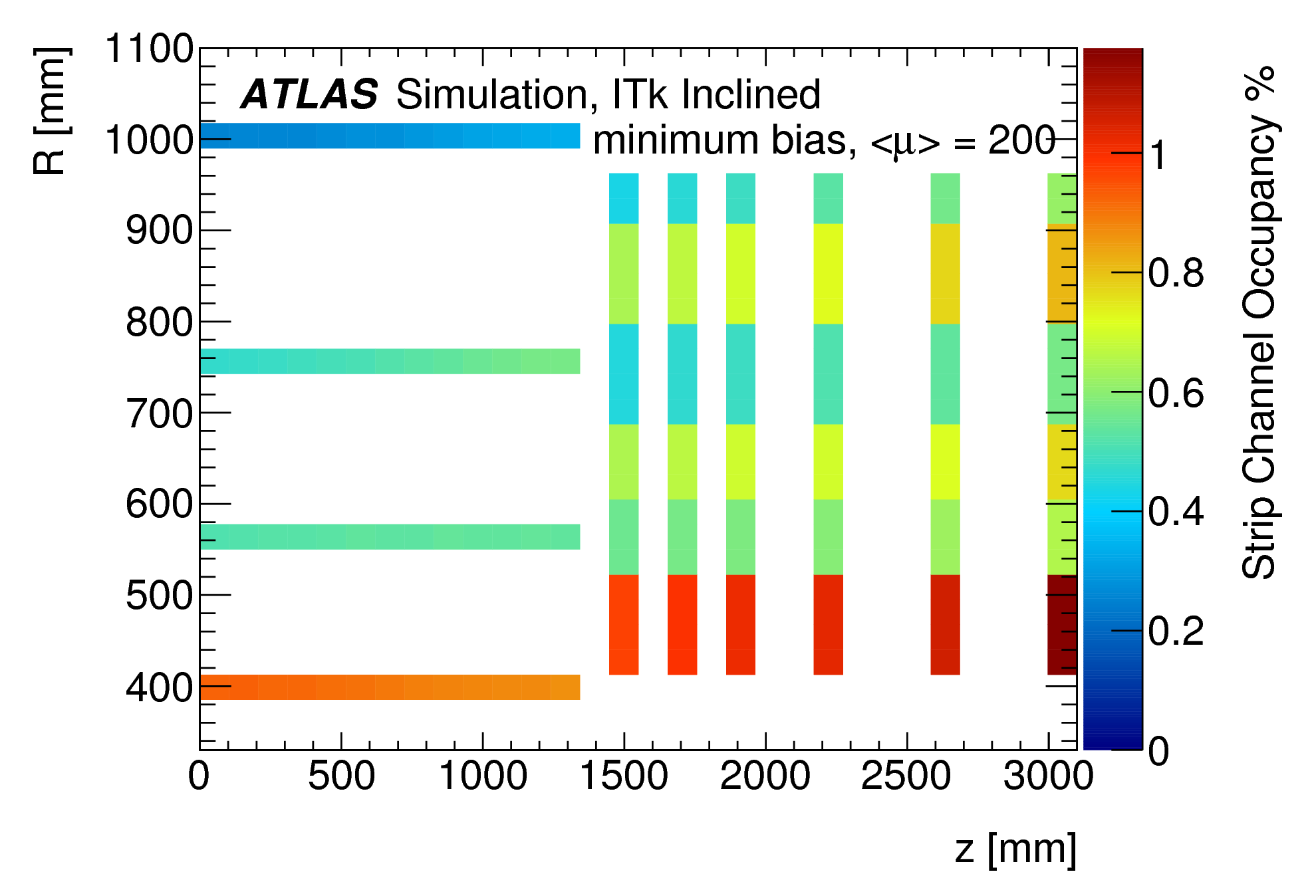}
\captionof{figure}{The average channel occupancy calculated with pileup=200 in the Strip tracker}
\label{fig:Occup}
\end{minipage}

\section{The new Strip detector structures}

Figure \ref{fig:sensor},\ref{fig:petal},\ref{fig:endcap} show the implementation of the new detector 
geometry structures into the ATLAS detector simulation framework; from left to right, the plots are corresponding 
to the endcap sensor, the endcap petal and and the whole endcap. 

\begin{minipage}[c]{.33\textwidth}
\includegraphics[width=1.0\textwidth]{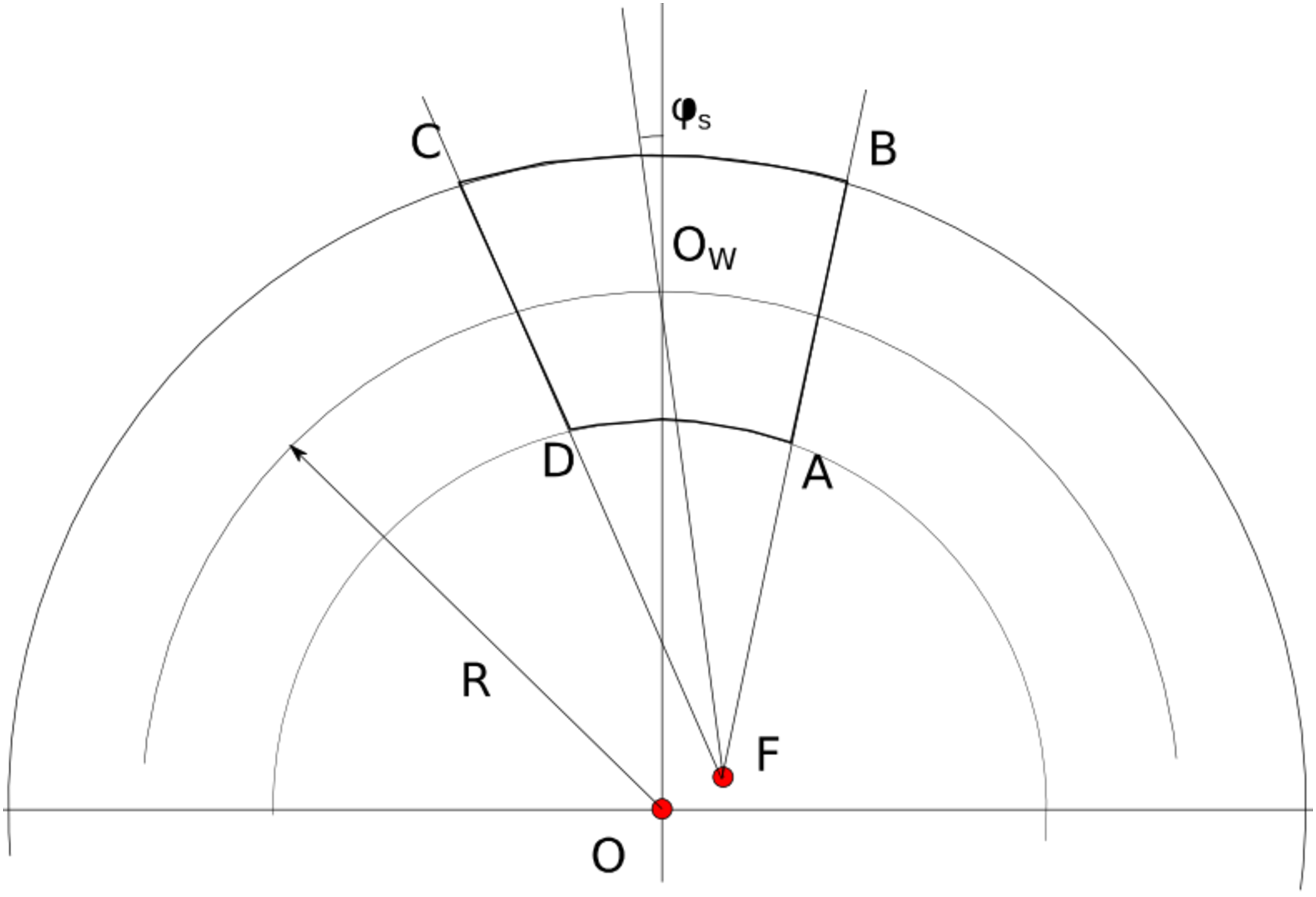}
\captionof{figure}{The stereo annulus ABCD is a endcap sensor}
\label{fig:sensor}
\end{minipage}
\hspace{0.5cm}
\begin{minipage}[c]{.33\textwidth}
\includegraphics[width=0.95\textwidth, height=0.2\textheight, angle=90,origin=c]{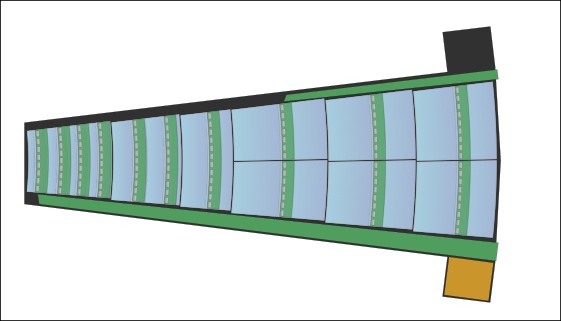}
\captionof{figure}{An endcap petal consisting of 6 rings and 9 sensors}
\label{fig:petal}
\end{minipage}
\hspace{0.1cm}
\begin{minipage}[c]{.33\textwidth}
\includegraphics[width=0.95\textwidth]{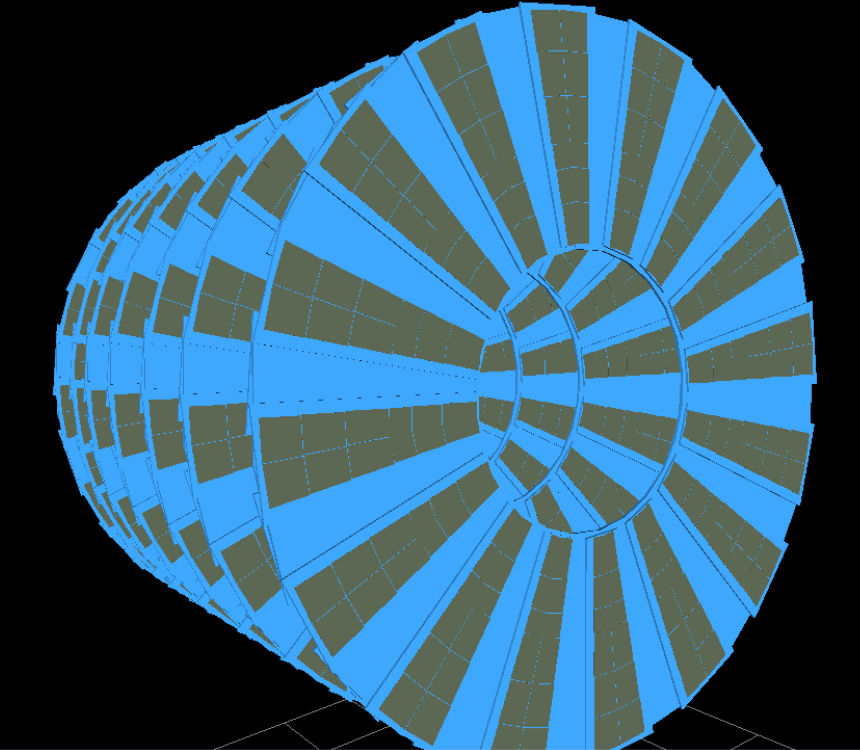}
\captionof{figure}{A whole endcap of the Strip detector}
\label{fig:endcap}
\end{minipage}

\section{The tracking performance}

Nice tracking performance are achieved for the future ITk, which could demonstrate
that the who software chain (including simulation as the first step) for the ITk has been done properly.    
From left to right, the plots are the mean number of hits per track as a function of $\eta$, 
the track reconstruction efficiency for particles as function of $\eta$ and
the resolutions on track parameters $z_0$, $q/p_T$ as a function of true track $\eta$.  

\begin{minipage}[c]{.493\textwidth}
\includegraphics[width=0.94\textwidth,height=0.2\textheight]{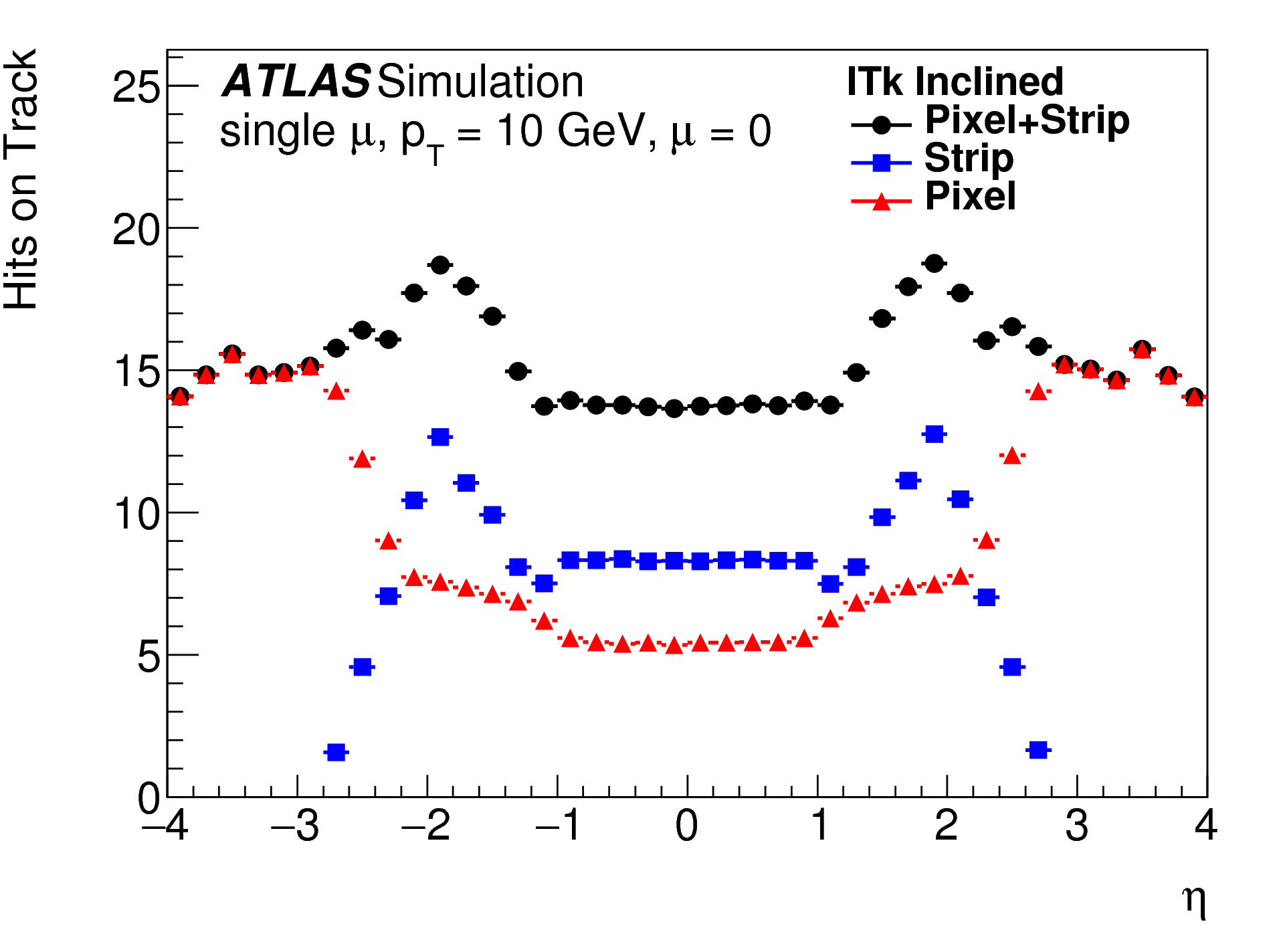}
\captionof{figure}{Number of hits per track vs. $\eta$}
\label{fig:trk1}
\end{minipage}
\begin{minipage}[c]{.493\textwidth}
\includegraphics[width=0.94\textwidth,height=0.2\textheight]{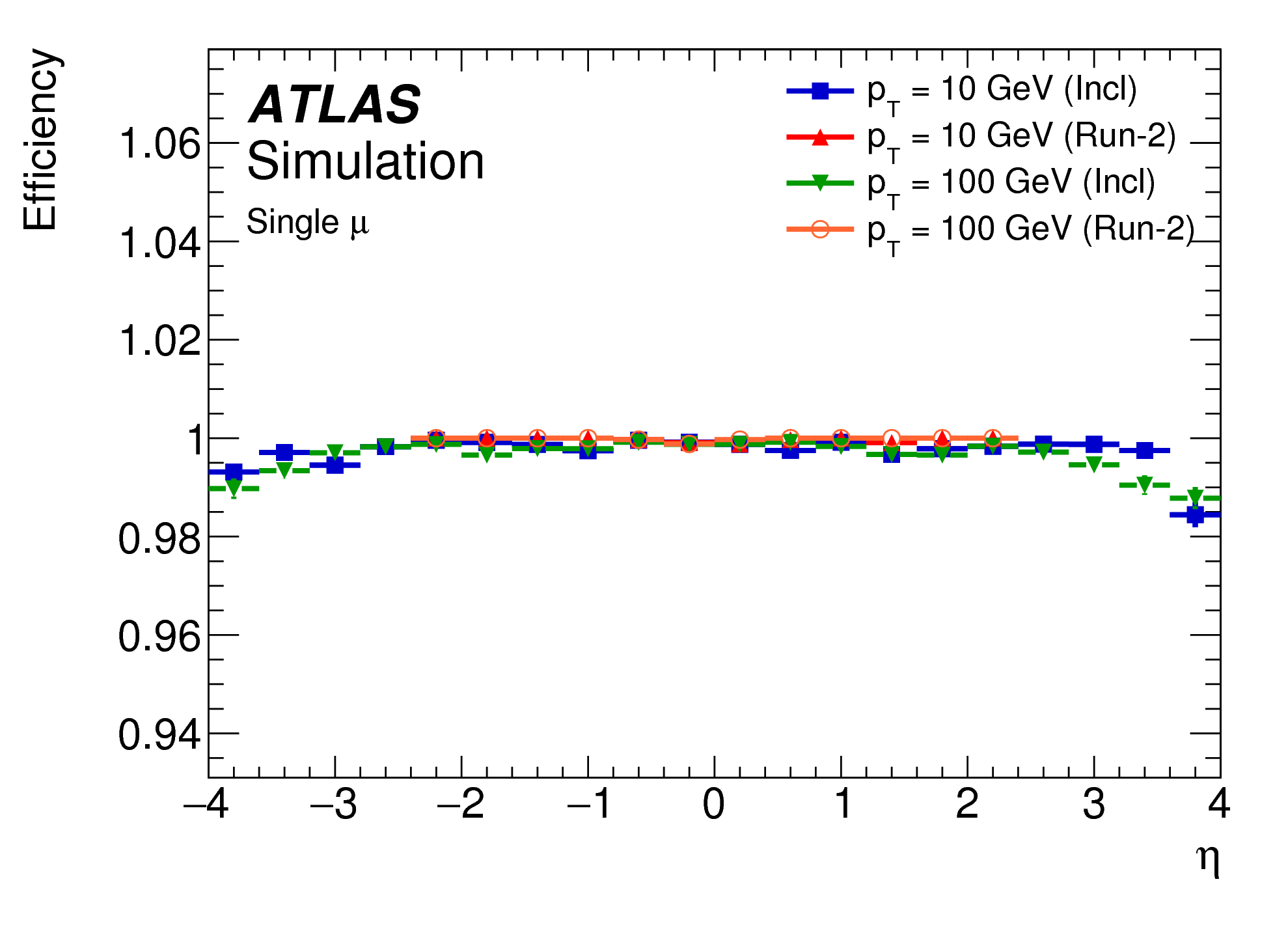}
\captionof{figure}{Track reconstruction efficiency vs. $\eta$}
\label{fig:trk2}
\end{minipage}  \\ 
\begin{minipage}[c]{.493\textwidth}
\includegraphics[width=1.0\textwidth,height=0.2\textheight]{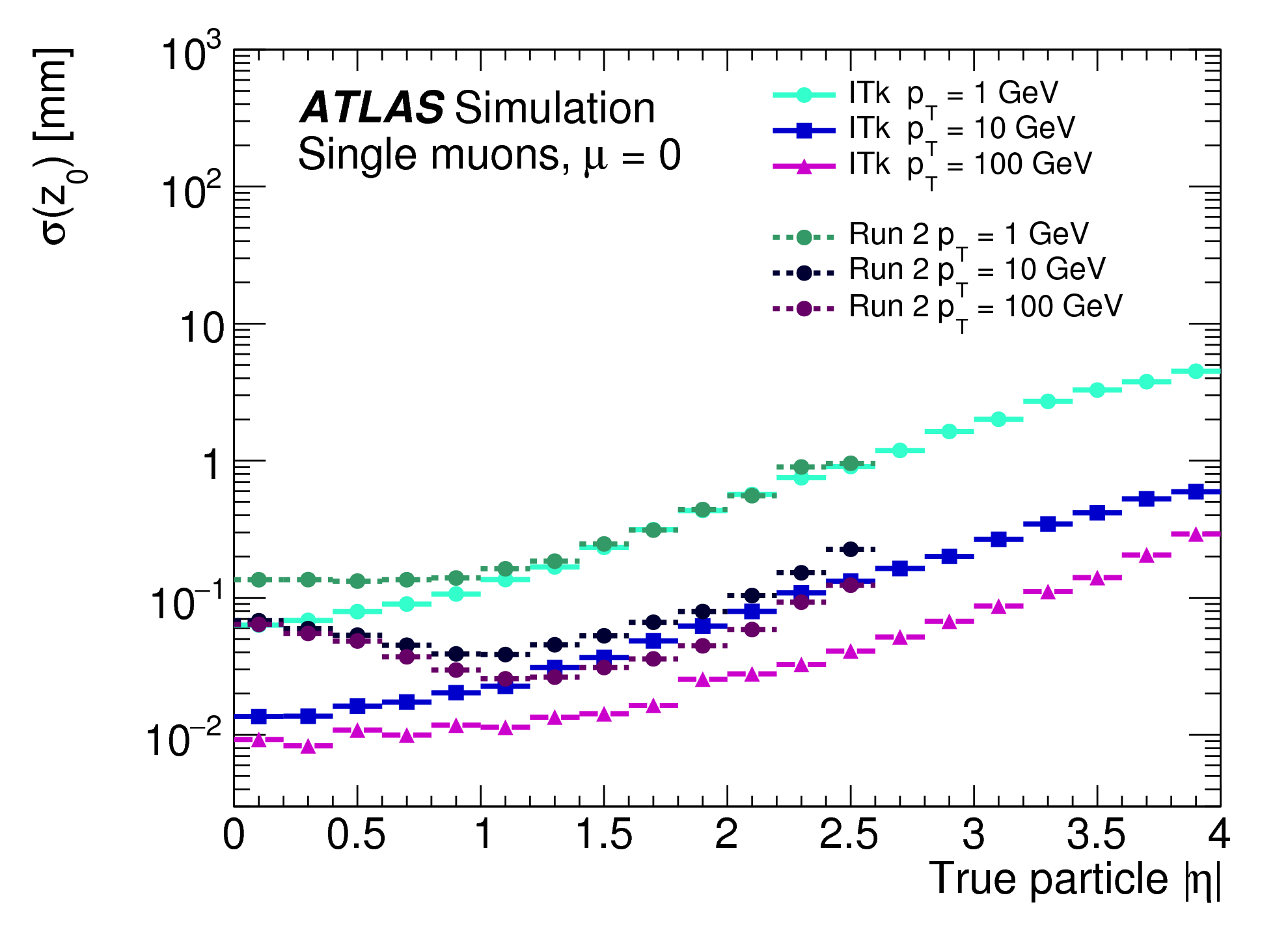}
\captionof{figure}{$z_0$ resolution vs. $\eta$}
\label{fig:trk3}
\end{minipage}
\begin{minipage}[c]{.493\textwidth}
\includegraphics[width=1.0\textwidth,height=0.2\textheight]{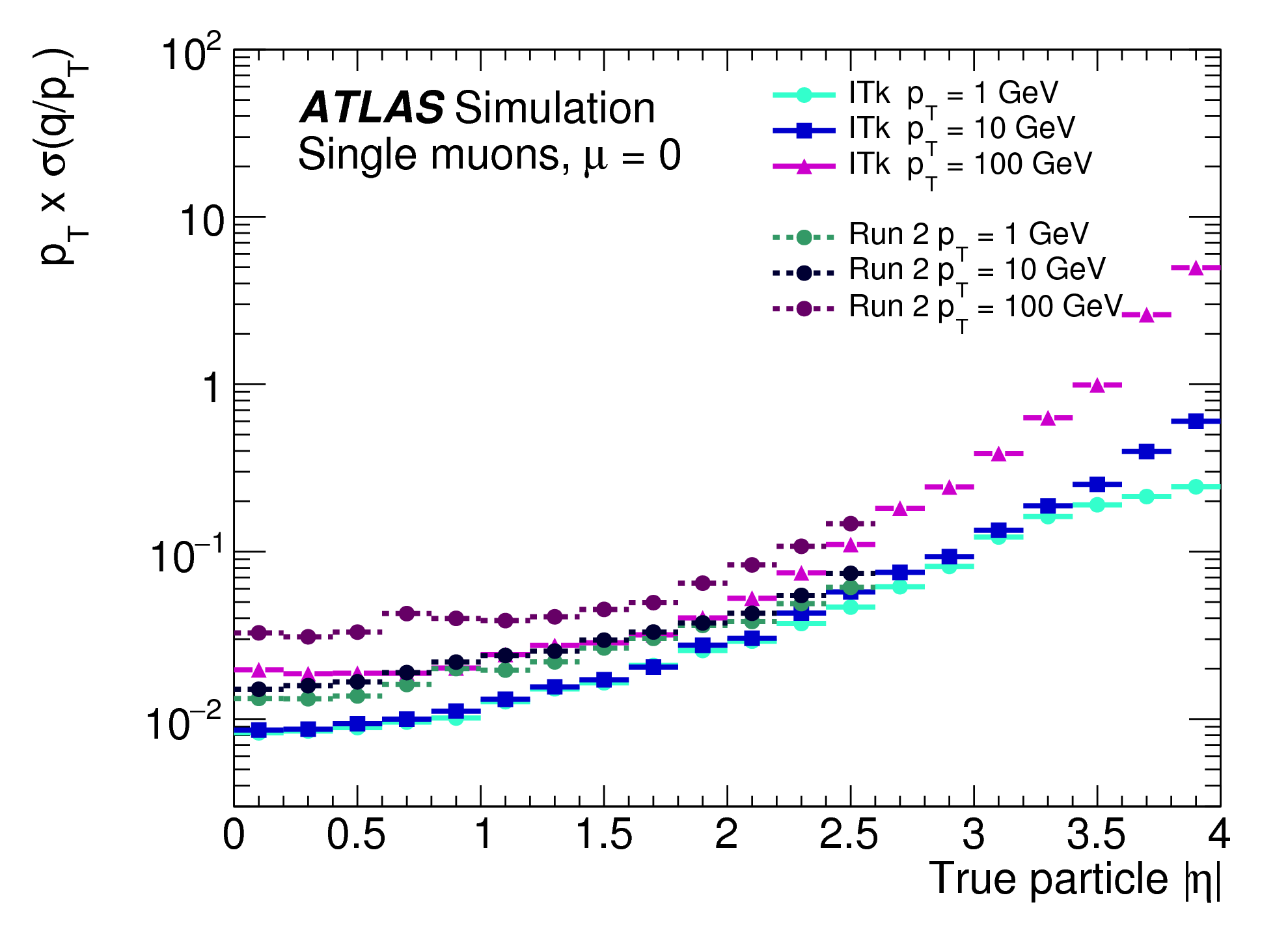}
\captionof{figure}{$q/p_T$ resolution vs. $\eta$}
\label{fig:trk4}
\end{minipage}


\section{Conclusions}
The extremely challenging situation at HL-LHC makes very hard to design the further new
tracker. Several years' work successfully converged on the simulation of ITk for the Strip Technical Design Report (TDR~\cite{Strip-TDR}), 
which was released in year 2017.

\end{document}